\documentclass{jetpl}
\twocolumn

\usepackage{epsfig,graphicx,amsfonts,amsmath,amssymb,bbm}

\RequirePackage{mathrsfs}
\usepackage{color}
\usepackage{epsf}
\usepackage{epstopdf}

\newcommand{\zz}[1]{{#1}}

\newcommand{\Le}{\left(}
\newcommand{\Ra}{\right)}

\newcommand{\beq}{\begin{equation}}
	\newcommand{\eeq}{\end{equation}}
\newcommand{\beqar}{\begin{eqnarray}}
	\newcommand{\eeqar}{\end{eqnarray}}
\usepackage{color}

\lat

\title{Riemann  - Cartan gravity with  dynamical signature }

\rtitle{Riemann  - Cartan gravity with  dynamical signature }

\sodtitle{Riemann  - Cartan gravity with  dynamical signature}

\author{S.Bondarenko$^{+}$, M.\,A.\,Zubkov $^{+}$\/\thanks{e-mail: mikhailzu@ariel.ac.il}}

\rauthor{S.\,Bondarenko, M.\,A.\,Zubkov }

\sodauthor{Zhang, Zubkov}

\address{$^+$ Physics Department, Ariel University, Ariel 40700, Israel}

\date{\today}

\abstract{Model of Riemann-Cartan gravity with varying signature of metric is considered. The basic dynamical variables of the formalism are vierbein, \zz{spin connection,} and an internal metric in the tangent space. The corresponding action contains new terms, which depend on these fields. In general case the signature of the metric is determined dynamically. The Minkowski signature is preferred dynamically because the configurations with the other signatures are dynamically suppressed. We also discuss briefly the motion of particles in the background of the modified black hole configuration, in which inside the horizon the signature is that of Euclidean space - time.  }

\begin{document}
	
	\maketitle

\section{Introduction}

We live in the four - dimensional world with Minkowski signature. From the mathematical point of view the signature $(1,-1,-1,-1)$ is only one of the possible choices. Therefore, the idea that the observed signature of space - time appears dynamically, attracts attention of theoretician working on both cosmology and quantum gravity  \cite{MisWh,MisWh_,Hawk1,Hawk1_,Hawk1__,Hawk1___,Hawk1____,Sakh,Ander,Ander_,Gero,Gero1,Gero_,Sork,Sork_,Strom,Dray,Dray_,Dray__,Dray___,Dray____,Dray_____,Viss,Barv,Barv_,Barv__,Barv___,Barv____,Barv_____,Barv______,Barv_______,borde,borde_,Kri,Kri_,SigC,SigC_,SigC__,SigC____,SigC____1,SigC_____,SigC______,SigC_______,SigC________,SigC_________,SigC__________,SigC___________,SigC____________,SigC2,Green,Green_,Spont}.  In principle, signature may be changed dynamically due to quantum fluctuations  \cite{MisWh}. Such fluctuations were  especially strong at the very beginning of the Universe \cite{Hawk1}. 

In the recent paper by one of the present authors \cite{B2021} the dynamical signature changes are realized within the systems defined on the manyfolds with complex - valued metric.  
The present paper is devoted to the more simple construction, in which the dynamical changes of signature are due to the real - valued metric $O_{ab}$ defined on tangent space. In case of the conventional Riemann - Cartan gravity such a metric is fixed and is equal to the one of Minkowski space - time. Now the fluctuations of matrix $O_{ab}$ give rise to the dynamical changes of the space - time signature. In addition to $O_{ab}$ the theory contains vierbein $e^a_\mu$ and spin connection. The latter belongs to the Lie algebra of the $SL(4,R)$ group that contains both Lorentz group and the groups $SO(4), SO(2,2)$.

 \section{Basic gravitational variables}

 Our first basic variable is vierbein $e_\mu^a$, which is matrix $4\times 4$. Metric is composed of vierbein as follows 
\beq\label{V18}
g_{\,\mu \nu}\,=\,O_{\,a b}\,e_\mu^ a\,e_\nu^b\,,
\eeq
the real symmetric matrix $O$ is our second dynamical variable, which plays the role of  metric on tangent space. \zz{Notice, that in our theory the metric is not the basic dynamical variable unlike conventional general relativity. Like in the Riemann - Cartan gravity tensor $g_{\,\mu \nu}$ is determined by Eq. (\ref{V18}) through the true dynamical variables. However, contrary to the Riemann - Cartan gravity in addition to the dynamical field $e_\mu^a$ there is an independent dynamical field $O_{ab}$.}

The case of space - time with Minkowski signature corresponds to the choice 
$
O={\rm diag}(1,-1,-1,-1)
$
while the case of Euclidean signature is
$
O={\rm diag}(1,1,1,1)\,.
$
The choices $O={\rm diag}(-1,1,1,1)$ and $O={\rm diag}(-1,-1,-1,-1)$ 
also represent Minkowski and Euclidean signatures correspondingly. The cases 
$
O={\rm diag}(-1,-1,1,1)
$
and 
$
O={\rm diag}(1,1,-1,-1)
$
represent the signature, which is typically not considered in the framework of conventional quantum field theory. $O(4)$ transformations $\Omega$ of vierbein 
$e^a_\mu \to \Omega_b^a e_\mu^b$ together with rescaling  $e^a_\mu \to \Lambda_b^a e_\mu^b$  (where $\Lambda = {\rm diag}(\lambda_1,\lambda_2,\lambda_3,\lambda_4)$ with positive $\lambda_i$)  are able to reduce the general form of matrix $O$ to the six above mentioned canonical forms.

In the theory with fixed matrix $O$ of one of the above mentioned canonical forms the vierbein belongs to representation of one of the three groups $SO(4), SO(3,1), SO(2,2)$. Correspondingly, the local gauge theory would contain the gauge field of one of the three Lie algebras. In our general case we need gauge field of the group, which contains $SO(4), SO(3,1), SO(2,2)$. 
A possible choice is $SL(4,R)$. Thus we introduce connection $\omega^a_{\mu b}$ that belongs to its algebra. Covariant derivative of the contravariant with respect to index $a$ vierbein receives the form
\beq\label{V24}
D_\mu e^a_\nu = (\partial_\mu \delta^a_b+ \omega^a_{\mu b })e^b_\nu
\eeq
while covariant derivative of internal metric $O$ with covariant indexes is:
\beq\label{V25}
D_\mu O_{ab} = (\delta_a^c \delta^d_b\partial_\mu - \omega^c_{\mu a }\delta^d_b- \omega^d_{\mu b}\delta^c_a)O_{cd}\,.
\eeq
As well as in conventional case we define the inverse vierbein matrices through
\beq\label{V26}
 \quad E_a^\mu e^a_\nu = \delta^\mu_\nu\,,\,\,\,\,\quad E_a^\mu e^b_\mu = \delta^b_a\,
\eeq
while metric with upper indices is defined through 
\beq\label{V27}
g^{\mu\nu}g_{\nu \rho} = E^\mu_a E^\nu_b {\cal O}^{ab} e_\nu^c e_\rho^d O_{cd} = \delta^\mu_\rho
\eeq
with ${\cal O}$ matrix such that
\beq\label{V28}
{\cal O}^{a b}O_{bc} =  \delta^a_c
\eeq

Notice that metric $O$ does not have discrete nature. The signature change may occur as a result of the smooth modification of the field $O$. Then $det O = 0$ at the transition hypersurface between the regions of space with different signatures. In the other words on such hypersurfaces metric is degenerate, and effectively the system has dimension smaller than $4$ along these hypersurfaces.  

\section{Dynamics of  fields}

One can construct the following action quadratic in the derivatives of $O$:
\beq\label{Act1}
S_O\,=\,
\int\, d^{4} x\,e\,\sqrt{O}\,
E^{\,\mu}_{c}\,E^{\,\nu}_{f}\,\Le D_{\mu} O \Ra_{\,a b}\,\Le D_{\nu} O \Ra_{d e}\,{\alpha}^{abc;def}
\eeq
with 
\begin{eqnarray}\label{Act11}
&& e\,=\,{\rm det}\, e = \frac{1}{4!} \, e^a_\mu e^b_\nu e^c_\rho e^d_\sigma \epsilon_{abcd} \epsilon^{\mu\nu\rho\sigma} \nonumber\\ && \sqrt{O}\,=\,\sqrt{{\rm det}\,O}\,\nonumber\\ && =\sqrt{\frac{1}{4!} \, O_{a_1 a_2} O_{b_1b_2} O_{c_1 c_2} O_{d_1 d_2} \epsilon^{a_1b_1c_1d_1} \epsilon^{a_2 b_2 c_2 d_2 }}.
\end{eqnarray}
Under the general coordinate transformation $x^\mu \to \tilde{x}^\nu$, $e_\mu^a \to \tilde{e}_\mu^a = e_\nu^a \frac{\partial {x}^\nu}{\partial \tilde{x}^\mu} $ the combination $d^{4} x\,e$ remains invariant:
$$
d^{4} x\,e \to d^{4} \tilde{x}\,\tilde{e}\, {\rm det}\,\Big( \frac{\partial x}{\partial \tilde{x}}\Big)\,{\rm det}\,\Big( \frac{\partial \tilde{x}}{\partial {x}}\Big) =  d^{4} \tilde{x}\,\tilde{e}
$$
At the same time under the $SL(4,R)$ transformation $e^a_\mu \to \Omega_b^a e_\mu^b = \tilde{e}^a_\mu$, $O_{ab} \to O_{cd} \tilde{\Omega}^c_a \tilde{\Omega}^d_b $  we have:
$$
e \,\sqrt{O} \to \tilde{e}\, {\rm det} \,\Big(\tilde{\Omega}\Big)\, \sqrt{\tilde{O}}\, \sqrt{{\rm det}^2 \,\Big({\Omega}\Big)}=\tilde{e}\, \sqrt{\tilde{O}}
$$
Therefore, Eq. (\ref{Act1}) remains invariant under both internal $SL(4,R)$ and space  - time diffeomorphisms.

Tensor $\alpha$ is to be composed of $O$. It may contain six different terms listed in Appendix \ref{AppendixA}.  
We may represent the above term in the action through the tetrad components of the derivatives of $O$:
$$
E^\mu_c D_\mu O_{ab} = O_{ab;c}
$$
With these notations we have
\beq\label{Act1_}
S_O\,=\,
\int\, d^{4} x\,e\,\sqrt{O}\,
O_{ab;c}\,O_{de;f}\,{\alpha}^{abc;def}
\eeq

Measure over $\cal O$ in the functional integral may be defined as the Haar measure associated with norm
$$
\|\delta {\cal O}\|^2 = \int d^4 x |e| \sqrt{| O|}  \delta {O}_{ab} \delta { O}_{cd} \Big(\beta_1{\cal O}^{ac}{\cal O}^{bd}+\beta_2 {\cal O}^{ab}{\cal O}^{cd}\Big)
$$
with real - valued coefficients $\beta_i$. These values should, of course, be taken in such a way that the above given norm is positive defined.

Modified Einstein-Cartan action reads 
\beq\label{Act3}
S_{\omega} = -m_P^2\int d^4 x\, e\, \sqrt{O}\, {\cal R}_{\mu \nu b}^a E^\mu_a E^\nu_d {\cal O}^{b d}
\eeq
Here $\cal R$ is curvature of gauge field $\omega$ defined as
\beq\label{Act4}
{\cal R}_{\mu \nu b}^a= \partial_\mu \omega^a_{b\nu} - \partial_\nu \omega^a_{b\mu} + \omega^a_{c\mu}\omega^{c}_{b\nu}-\omega^a_{c\nu}\omega^{c}_{b\mu}
\eeq
Besides, we may consider terms quadratic in curvature.  In order to classify these terms we introduce first the tetrad components of curvature:
$$
{\cal R}_{abcd} = E^\mu_c E^\nu_d O_{ad} {\cal R}^d_{ \mu \nu b}
$$
The general form of the action quadratic in curvature has the form: 
\beq\label{Act5_}
S_R =  \int d^4 x \, e\, \sqrt{O}\, {\cal R}_{a_1b_1c_1d_1} {\cal R}_{a_2 b_2 c_2 d_2} \gamma^{a_1b_1c_1d_1a_2b_2c_2d_2}\,,
\eeq
Tensor $\gamma$ is composed of matrices ${\cal O}$ and is discussed in Appendix \ref{AppendixA}. Unlike the case of the conventional gravity in space - time of Euclidean signature (or Minkowski signature) there are much more terms squared in curvature. The reason is the absence of symmetry with respect to the interchange of indexes $a,b$ in ${\cal R}_{abcd}$.
 
Another terms in the action may be composed of the covariant derivative of vielbein
$$
e^a_{\mu ;\nu} = D_\nu e^a_\mu
$$
In particular, tensor of torsion is defined as
\beq\label{Act6}
T^a_{\mu\nu} = D_\mu e^a_\nu - D_\nu e^a_\mu  
\eeq
We define the tetrad components of the derivatives of vierbein as
$$
e_{a b ;c} = O_{ad}E^\mu_b E^\nu_c D_\mu e^d_\nu
$$
There may be several independent terms quadratic in this derivative. Those ones have the form
\beq\label{Act7_}
S_e = \int d^4 x\, e\, \sqrt{ O}\, e_{a_1 b_1;c_1} e_{a_2 b_2;c_2} \zeta^{a_1 b_1 c_1 a_2 b_2 c_2}
\eeq
The most general form of tensor $\zeta$ may also be read off from Appendix \ref{AppendixA}.
Besides, we refer here to \cite{Diakonov}, where the classification of the similar terms has been given for the case of ordinary Riemann - Cartan gravity.

 There is also the mixed term 
\beq\label{Act7}
S_{Oe} = \int d^4 x\, e\, \sqrt{ O}\, e_{a_1 b_1; c_1} O_{a_2 b_2; c_2} \eta^{a_1 b_1 c_1 a_2 b_2 c_2}
\eeq
with parameters $\eta^{a_1 b_1 c_1 a_2 b_2 c_2}$ that are as well discussed in Appendix \ref{AppendixA}. Such terms do not present in the constructions of \cite{Diakonov}.
 Finally, one may add the trivial cosmological constant term:
\beq\label{Act8}
S_\lambda = -\lambda \int d^4 x\, e\, \sqrt{ O}\,,
\eeq

Partition function may be written as
\begin{eqnarray}
	Z= \int D e D O D \omega e^{-S_O  -S_\omega -S_R- S_e - S_{Oe}- S_\lambda}\label{Z}
\end{eqnarray}
One can always choose the coefficients in the action ($\alpha_i,  \zeta_\sigma, \gamma_\sigma, \eta_\sigma, \lambda$) in such a way that the action is bounded from below for the case of real positive $\sqrt{{\rm det} O}$. Moreover, we require that Euclidean action is positively defined. This allows to define the self - consistent quantum theory. In this theory the fluctuations of fields appear to be exponentially suppressed for real positive $\sqrt{{\rm det} O}$. At the same time negative ${\rm det} O$ results in the appearance of imaginary unity in the exponent of Eq. (\ref{Z}). The corresponding configurations are not exponentially suppressed and dominate over the configurations with positive  ${\rm det} O$. {\bf This is the way how the signature 
$(1,-1,-1,-1)$ (or $(-1,1,1,1)$) is distinguished dynamically.}

For the completeness let us notice here that matter fields in the given theory should belong to the irreducible representations of group $SL(4,R)$. Those represenations are much more complicated than the conventional representations of the $SO(3,1)$ and $SO(4)$ groups. For their explicit construction see \cite{Spinors1,Spinors2} and references therein. Dirac - like action for the corresponding spinor fields may be given in terms of these spinors \cite{Spinors3,Spinors4}. More specifically, the corresponding Dirac equation is given by Eq. (44) of \cite{Spinors4}. In the domains of the theory, where the signature is fixed, the physical fields are reduced to those of the irreducible representations of the groups $SO(3,1), SO(4), SO(2,2)$. 


\section{Modified black hole configuration with signature change over the horizon}\label{Shell} 

As an illustration of the proposed theory  let us consider the configuration that looks like the ordinary neutral non - rotating black hole outside of the horizon but appears to have the Euclidean signature inside the horizon. Here we assume also that torsion is vanishing, i.e. the spin connection is expressed unambigously through metric. Notice that we do not mean here that the considered configuration is a solution of classical field equations for gravitational field. Instead we assume here that this is a fixed configuration that serves as a background for motion of a massive particle.  

In the Painleve – Gullstrand reference frame metric of the black hole outside of the horizon has the form
\begin{equation}
	ds^2 = dt^2 - (d{\bf r} - {\bf v}({\bf r}) dt)^2, \label{PGQ}
\end{equation}
where
\begin{equation}
	{\bf v} = -\frac{1}{m_P} \frac{\bf r}{r}\, \sqrt{\frac{2M}{r}} = -\frac{\bf r}{r} v(r)
\end{equation}
is the velocity of the free falling body (measured in its self time).
Here  $m_p$ is the Plank mass, $M$ is the mass of the black hole. The position of the horizon corresponds to $r_c = 2M/m_P^2$.

The given form of metric corresponds to the following expressions for  $E^\mu_a$ and $e_\mu^a$:
\begin{equation}
	E^\mu_a = \left(\begin{array}{cc}1 & {\bf v}\\0 & 1\end{array} \right)
\end{equation}
and 
\begin{equation}
	e_\mu^a = \left(\begin{array}{cc}1 & -{\bf v}\\0 & 1\end{array} \right)
\end{equation}
We assume that those expressions remain valid also inside the horizon.
The only nonzero component of the spin connection $\omega_{ab \mu}$ corresponds to the space - time index $\mu$ of radial direction, the indices $a,b$ corresponding to time direction and radial direction. Outside of the horizon it has the form
\begin{equation}
	\omega_{r0r} = -\omega_{0rr} = -\frac{d}{dr}|{\bf v}({\bf r})|\label{omega}
\end{equation}

Vacuum is falling with the velocity ${\bf v}({\bf r})$ defined above.  At any given point of space - time in the accompanying reference frame scalar particles/anti - particles have  dispersion
$$ E(p) = \pm \sqrt{p^2 + m^2}$$
Its upper branch is empty while the lower branch is occupied. However, when we  look at this pattern from the Painleve -- Gullstrand reference frame, the dispersion  is 
$$ E({\bf p}) = \pm \sqrt{p^2 + m^2} + {\bf v} {\bf p}$$
Inside the horizon the Dirac cone (it appears at $p \to \infty$) would be overtilted in the true black hole (BH) configuration. In our case inside of the horizon, at $r < r_c$ 
 metric has the form:
 \begin{equation}
 	ds^2 = -dt^2 - (d{\bf r} - {\bf v}({\bf r}) dt)^2, \label{PGQ_}
 \end{equation}
This corresponds to matrix $O$ of the form
\begin{equation}
O = \Big( \begin{array}{cc} {\rm sign}(r-r_c)& 0 \\ 0 & - 1 \end{array}\Big)\label{O}
\end{equation}
Namely, inside the horizon the signature is $(-1,-1,-1,-1)$, i.e. that of Euclidean space - time. \zz{It is worth mentioning that the given configuration of the gravitational variables represents the toy model of the black - hole like system with the signature change. Such a configuration (or a configuration that is close to it) may actually appear as a classical solution of equations of motion following from the action of Eq. (\ref{Z}) for a certain choice of parameters listed in Appendix \ref{AppendixA}. However, the explicit construction of the corresponding action that gives rise to classical solution having the form of Eq. (\ref{O}) remains out of the scope of the present paper.}

Below in this section we consider the motion of particles neglecting their internal $SL(4,R)$ structure.  

Inside the horizon there is no notion of time, but the particles exist in the sence that these are the point - like objects that have trajectories in Euclidean space - time - the lines that may be parametrized as $\bar{r}(t)$ (the three dimensional vector depending on the fourth coordinate denoted here by $t$). We describe the corresponding classical dynamics in Appendix \ref{AppendixB}. One can see that the notion of energy may be defined here as well as in Minkowski space - time, and the energy remains constant along the classical trajectory. 

Correspondingly, outside of the horizon the classical trajectory corresponding to energy $E$ carries action 
$$
S = \int \Big(-E dt + \frac{Ev(r) \pm \sqrt{E^2 - m^2(1-v^2(r))}}{1-v^2(r)}dr\Big)
$$
Its contribution to the partition function is $e^{i S}$. 
If classical particle of energy $E$ falls down to the center of the black hole, it might be reflected from the horizon that separates space - time pieces with different signatures. Also it might   undergo the transition through the horizon. If a particle had energy $E$ outside of the horizon, it acqures  Euclidean "energy" $E^\prime$ behind it. In the present paper we do not describe quantitatively energy change $E^\prime - E$ accross the horizon and do not calculate the probabilities of reflection and transmission. Such a description would require more precise determination of physics at the interphase between the Euclidean and Minkowski domains. Recall that for $E\ge m$ the particle may approach to the horizon from infinity, while for $E>m$ particle might exist classically within a finite region outside of the horizon. 

It is worth mentioning, that inside the horizon the classical particles are able to approach the horizon unlike the case of the ordinary black hole. If those particles escape from the interior of the horizon, then they get infinite momentum. This points out that such a process is purely quantum. In the case of space - time with Minkowski signature the infinite value of momentum on the horizon for the outgoing particles is also a consequence of the impossibility for the classical particles to escape. Those that fall down to the center of the BH do not have infinite momentum when pass through the horizon.

We do not describe here the quantum process, in which particles escape from the interior of the horizon. One of the reasons for this is that in a more realistic situation matrix $O$ is assumed not to be changed in a step - like way. The signature change is expected to be acompanied by vanishing of the determinant of $O$, which results from smooth variation of $O$. This would bring us to the description of a kind of a domain wall between the two pieces of space - time. These domain walls have finite thickness, and the dynamics of particles while they traverse them might be very complicated. It also depends on the particular form of the dependence of $O$ on $r$. The description of such  domain walls is out of the scope of the present paper.  
  
The Euclidean action in terms of energy $E^\prime$ along the classical trajectory inside the BH is given by (See Appendix \ref{AppendixB}) 
$$
S_E = \int \Big(E^\prime dt + \frac{-E^\prime v(r) \pm \sqrt{-(E^\prime)^2 + m^2(1+v^2(r))}}{1+v^2(r)}dr\Big)
$$
This trajectory contributes the semi - classical expression for the partition function with the factor $e^{-S_E}$. This factor suppresses the presence of particles in the interior of the BH. However, those ones that fall down to the interior of the BH proceed falling down (being considered on the classical level). 

We conclude that quantum dynamics of  particles falling down to the BH results in reflection from the horizon. There is also the transmission probability to the interior. Such a configuration may be distinguished from the ordinary BH through the observation of the mentioned above reflection and also through the observation of the radiation from its interior. We expect that such radiation differs from the Hawing one. 

One may also consider the BH - like configuration, in which the signature change (change of matrix $O$) occurs somewhere at $r_c^\prime < r_c$, not precisely at the position of the horizon. Such a configuration is essentially different from the one considered above. Namely, the radiation from the region of its interior adjacent to the horizon is the Hawking one. Besides, there is no reflection from the horizon. Instead, reflection occurs from the surface with $r = r_c^\prime$, but those particles remain inside the BH (being considered classically).

 \section{Conclusion}

To conclude, in the given paper we propose the relatively simple theory that admits  the dynamical changes of space - time signature. We classify all possible terms in the corresponding gravitational action  containing up to the second derivatives of basic variables. We argue that certain choice of the coupling constants leads to the dynamical suppression of the signatures other than the Minkowski signature. This follows from the fact that the considered terms in the action contain the square root of the determinant of internal metric $O_{ab}$. This square root is imaginary for the case of Minkowski signature and is real for the other cases. As a result the exponent entering the partition function has the form $e^{i |S|}$ for space - time of Minkowski signature, and the form $e^{-S}$ for the other cases. Then the positively defined action $S$ results in the dynamical suppression of space - time with alternative signatures.

As an illustration of our general construction we considered roughly the modification of the black hole solution, in which outside of the horizon it looks like an ordinary Schwarzshield solution (considered in Gullstrand - Painleve reference frame). Inside of the horizon the expression for the vielbein remains the same as in the ordinary Painleve - Gullstrand black hole, but the matrix $O_{ab}$ changes signature to that of Euclidean space - time. We do not have an intention to consider the given configuration as a real classical solution, buth rather look at it as a toy model of the black hole - like configuration with the signature change. On the background of this configuration the motion of a massive particle is considered briefly. 
It is worth mentioning that such a configuration may appear at a certain stage of the gravitational collapse, when the singularity appears in the classical solution of Einstein equations close to center of the BH. Then the quantum dynamics comes into play, and the signature change might occur inside the horizon.

\appendix

\section{Coupling constants}
\label{AppendixA}

Here we represent the tensors of coupling constants encountered in the main text. These tensors are expressed through the components of ${\cal O}^{ab}$ as follows.  
\begin{eqnarray}\label{Act2}
&&{\alpha}^{abc;def} = \alpha_1 {\cal O}^{ab}{\cal O}^{de}{\cal O}^{cf} + \alpha_2 {\cal O}^{ac}{\cal O}^{df}{\cal O}^{be} + \alpha_3 {\cal O}^{ab}{\cal O}^{df}{\cal O}^{ce} \nonumber\\&& + \alpha_4 {\cal O}^{ad}{\cal O}^{be}{\cal O}^{cf} + \alpha_5 {\cal O}^{af}{\cal O}^{bd}{\cal O}^{ce} + \alpha_6 {\cal O}^{ac}{\cal O}^{df}{\cal O}^{be}
\end{eqnarray}
Here $\alpha_i$ are real - valued constants. Tensor $\gamma^{a_1 a_2 a_3 a_4 a_5 a_6 a_7 a_8}$ contains much more independent terms composed of $\cal O$. We may represent it as
\begin{eqnarray}
&&\gamma^{a_1 a_2 a_3 a_4 a_5 a_6 a_7 a_8} =\label{GB} \\&&=\sum_{\sigma} \gamma_{\sigma}{\cal O}^{a_{\sigma(1)} a_{\sigma(2)}}{\cal O}^{a_{\sigma(3)} a_{\sigma(4)}}{\cal O}^{a_{\sigma(5)} a_{\sigma(6)}}{\cal O}^{a_{\sigma(7)} a_{\sigma(8)}}\nonumber
\end{eqnarray}
Constants $\gamma_\sigma$ are real - valued.
Here the sum is over $\frac{1}{4!}C_8^2 C_6^2 C_4^2 = \frac{8!}{4! 2^4} = 105$ permutations $\sigma(i)$ of $8$ integer numbers $1,2,...,8$ that give rise to four different pairings between them. Some of these terms are irrelevant since the corresponding contributions to ${\cal R}_{a_1b_1c_1d_1} {\cal R}_{a_2 b_2 c_2 d_2} \gamma^{a_1b_1c_1d_1a_2b_2c_2d_2}$ vanish identically because of the symmetry  
$$
{\cal R}_{abcd}= - {\cal R}_{abdc}
$$
Namely, there are $C_6^2 C_4^2/3! + C_6^2 C_4^2/3! - C_4^2/2! = 27$ such terms. 
Notice that the remaining $78$ terms in Eq. (\ref{GB}) are not all independent because some of them give contributions to ${\cal R}_{a_1b_1c_1d_1} {\cal R}_{a_2 b_2 c_2 d_2} \gamma^{a_1b_1c_1d_1a_2b_2c_2d_2}$ proportional to each other.  
For brevity we do not represent here the complete set of relevant independent terms entering Eq. (\ref{GB}). 

The similar situation takes place for tensor
$\zeta^{a_1 b_1 c_1 a_2 b_2 c_2}$. Namely, it may be represented as 
\begin{eqnarray}
	&&\zeta^{a_1 a_2 a_3 a_4 a_5 a_6} =\label{ZB} \\&&=\sum_{\sigma} \zeta_{\sigma}{\cal O}^{a_{\sigma(1)} a_{\sigma(2)}}{\cal O}^{a_{\sigma(3)} a_{\sigma(4)}}{\cal O}^{a_{\sigma(5)} a_{\sigma(6)}}\nonumber
\end{eqnarray}
There are $C_6^2 C_4^2/3! = 15$ terms and $15$ real - valued constants $\zeta_\sigma$ in this expression that correspond to $3$ different pairings of $6$ integer numbers. Again, some of them give the contributions to  $e_{a_1 b_1;c_1} e_{a_2 b_2;c_2} \zeta^{a_1 b_1 c_1 a_2 b_2 c_2}$ proportional to each other. 

Tensor $\eta^{a_1 b_1 c_1 a_2 b_2 c_2}$
entering combination $e_{a_1 b_1; c_1} O_{a_2 b_2; c_2} \eta^{a_1 b_1 c_1 a_2 b_2 c_2}$ may be represented as 
\begin{eqnarray}
	&&\eta^{a_1 a_2 a_3 a_4 a_5 a_6} =\label{EB} \\&&=\sum_{\sigma} \eta_{\sigma}{\cal O}^{a_{\sigma(1)} a_{\sigma(2)}}{\cal O}^{a_{\sigma(3)} a_{\sigma(4)}}{\cal O}^{a_{\sigma(5)} a_{\sigma(6)}}\nonumber
\end{eqnarray}
Not all of these terms are relevant because of the symmetry $O_{a b; c} = O_{b a; c}$. Among the $15$ different real valued constants  
 $\eta_\sigma$ there are those that give contributions to $e_{a_1 b_1; c_1} O_{a_2 b_2; c_2} \eta^{a_1 b_1 c_1 a_2 b_2 c_2}$ proportional to each other. 
 
 We do not solve here completely the combinatorial problem about the number of independent constants  $\gamma_\sigma$, $\zeta_\sigma$,  $\eta_\sigma$. However, the reader can see already from the above consideration that there are of the order of $50$ such  independent constants, which results in a large variety of possible  gravitational actions. 

\section{Classical dynamics of particle in Euclidean space - time}
\label{AppendixB}

Let us consider for simplicity the radial motion of particle (we neglect dynamics of internal degrees of freedom). Action for the particle  has the form:
\begin{equation}
	S = m \int \sqrt{g_{\mu\nu}(x) dx^\mu dx^\nu}= i S_E
\end{equation} 
where $S_E$ is the so - called Euclidean action. (Here we consider the case when metric signature is $(-1,-1,-1,-1)$.) 
Each particle trajectory contributes functional integral $Z = \int Dx(t) e^{iS[x]} = \int Dx(t) e^{-S_E[x]} $ with weight
$e^{-S_E}$. The dominant contribution to  the one - particle partition function is given by a semi - classical trajectory with $\delta S_E = 0$. It is easy to rewrite the partition function in this semi - classical approximation through the functional integral over the trajectories in phase space (as it was noticed above we restrict ourselves by consideration of radial motion with radial coordinate $r$ and the corresponding momentum $p$):
$$
Z = \int Dr(t) Dp(t) e^{- \int (H(p(t),r)dt + p(t) dr(t))}
$$
with 
$$
H(p) = v(r) p  + \sqrt{m^2 - p^2}
$$
The Euclidean Hamilton equations follow from the variational principle 
$$
\delta \int (H(p(t),r)dt + p(t) dr(t)) = 0
$$
It gives 
\begin{eqnarray}
	\partial_p H & = & - \dot{r} \label{H1}\\
	\partial_r H &=& \dot{p}
\end{eqnarray}
Using Eq. (\ref{H1}) we express momentum $p$ through $\dot{q}$ and $q$. This gives
$$
Z = \int Dr(t)  e^{- m \int dt \sqrt{1 + (v + \dot{r})^2}} 
$$
The above mentioned Euclidean Hamiltonian equations also result (as well as in the case of Minkowski space - time) in the constant value of function $H(p(t),r(t))$ along the classical trajectory (i.e. along the trajectory determined by the Hamilton equations): $dH(p(t),r(t))/dt=0$. We, therefore, as in case of Minkowsli signature define the notion of energy $E = H(p(t),r(t))$ as the quantity that remains constant along each classical trajectory. Written in terms of $E$ the action along the classical trajectory is given by 
$$
S_E = \int \Big(E dt + \frac{-Ev(r) \pm \sqrt{-E^2 + m^2(1+v^2(r))}}{1+v^2(r)}dr\Big)
$$
The given trajectory contributes the semi - classical expression for the partition function with the factor $e^{-S_E}$.

\end{document}